\documentclass{mem}
\usepackage{natbib}\usepackage{txfonts}\usepackage{balance}
\usepackage{graphicx}
\usepackage[breaklinks,dvipdfm]{hyperref}
\idline{83}{0}
\def \th {\thinspace}

\def\approxgt{\mathrel{\hbox{\rlap{\lower.55ex \hbox {$\sim$}} \kern-.3em \raise.4ex \hbox{$>$}}}}
\def\lesssim{\mathrel{\hbox{\rlap{\lower.55ex \hbox {$\sim$}} \kern-.3em \raise.4ex \hbox{$<$}}}}
\def\approxlt{\mathrel{\hbox{\rlap{\lower.55ex \hbox {$\sim$}} \kern-.3em \raise.4ex \hbox{$<$}}}}
\def \degmark {^\circ}

\def \NH{$N_{\rm H}$}
\def \cygx2 {\hbox Cyg\thinspace X-2}
\def \gx5 {\hbox GX\thinspace 5-1}
\def \gx340 {\hbox GX\thinspace340+0}

\begin{document}

\title{
Dipping - versus Flaring in Z-track sources: resolving the controversy
}

\subtitle{}

\author{
M. Ba\l uci\'nska-Church\inst{1}
M. J. Church\inst{1}
\and A. Gibiec\inst{2}}

\offprints{M. Ba\l uci\'nska-Church}

\institute{
School of Physics and Astronomy, University of Birmingham,
Birmingham B15 2TT, U.K.
\and
Astronomical Observatory, Jagiellonian University, ul. Orla 171, 
30-244 Cracow, Poland
\email{mbc@star.sr.bham.ac.uk}
}

\authorrunning{Ba\l uci\'nska-Church }
\titlerunning{Dipping versus Flaring}

\abstract{
We review the longterm confusion which has existed over the nature of
flaring in the brightest
class of low mass X-ray binary: the Z-track sources, specifically in the 
Cygnus\th X-2 sub-group. Intensity reductions in the lightcurve produce a 
branch in colour-colour diagrams similar to that of real flares in the Sco\th 
X-1 like group, and the nature of this branch was not clear. However, based on 
observations of Cygnus\th X-2 in which this dipping/flaring occurred it was 
proposed that the mass accretion rate in Z-track sources in general increases 
monotonically along the Z-track towards the Flaring Branch, a standard 
assumption widely held. It was also suggested that the Cygnus\th X-2 group
have high inclination. Based on recent multi-wavelength observations of 
Cygnus\th X-2 we resolve these issues, showing by spectral analysis that the 
Dipping Branch consists of absorption events in the outer disk, unrelated to 
the occasional real flaring in the source. Thus motivation for $\dot M$ 
increasing along the Z from Horizontal - Normal to Flaring Branch is 
removed, as is the idea that high inclination 
distinguishes the Cygnus\th X-2 group. Finally, the observations provide 
further evidence for the extended nature of the Accretion Disk Corona (ADC), 
and the correct modelling of the ADC Comptonized emission is crucial to the 
interpretation of low mass X-ray binary data.
}
\maketitle{}

\section{Introduction}

The Z-track sources are the most luminous low mass X-ray binaries emitting at and above the Eddington
limit. Spectral variability presented in a hardness-intensity or colour-colour
diagram shows three distinctive tracks having a Z-shape 
known as the Normal (NB), Horizontal (HB) and Flaring (FB) Branch 
(Hasinger \& van 
der Klis 1989). This suggests major physical differences at the inner disk and 
the neutron star. There are six main Galactic Z-track sources known:  Cyg\th X-2, 
GX\th 340+O, GX\th 5-1, Sco\th X-1, GX\th 17+2, GX\th 349+2) and also the
transient source (XTE\th J1701-462) (Lin et al. 2009). Based on the shapes,
hardness-intensity diagrams of the Z-track sources are divided 
into two groups: the Cyg-like sources with a short Flaring Branch and a 
distinct Horizontal Branch, and the Sco-like sources with a strong Flaring
Branch and a weak Horizontal Branch. The nature of the three states and the cause of
the differences between the two sub-groups have not been understood.

In this paper we will concentrate on the longterm confusion between absorption dipping
and flaring found in the Cygnus\th X-2 like Z-track sources. Dipping in Cygnus\th X-2 was
first detected by Bonnet-Bidaud \& van der Klis (1982). In the pioneering work
of Hasinger and co-workers the spectral and timing properties of the Z-track sources
were revealed. Hasinger \& van der Klis (1989) showed that Cyg\th X-2 and GX\th 340+0
exhibited full Z-tracks in colour-colour diagrams with three branches. However,
the Flaring Branch, i.e. the lower branch on the Z as seen in other Z-track sources,
was associated with intensity {\it decreases}. Hasinger et al. (1990) in a multi-wavelength campaign 
on Cygnus\th X-2 using {\it Ginga} similarly found that the there were intensity decreases on the FB 
that ``could be mistaken for absorption dips''.  Kuulkers \& van der Klis (1995) argued that in 
\hbox{Cyg\th X-2} and GX\th 340+0 the FB in colour-colour representations corresponds to X-ray dipping 
(while in the Sco-like source the FB corresponds to strong flaring increases of intensity). Hence they 
suggested that the Cyg-like sources have high inclination angle distinguishing them from the Sco-like 
sources and proposed a model for the Cyg-like Sco-like differences involving absorption or scattering in 
the inner disk. Previously, there has been a lack of reliable spectral fitting of the Dipping Branch 
to establish its nature. In the present work, we carry out systematic spectral fitting
of the Dipping Branch using data from our 2009 multi-wavelength campaign (Ba\l uci\'nska, M., Schulz, N., 
Wilms, J., Gibiec, A., Hanke, M., Spencer, R.\ E., Rushton, A., Church, M.\ J., 2011, A\&A, 530, A102) 
and show that 
the dip events are unrelated to flaring. Thus there is now no motivation for an explanation 
of the Z-track sources based on inclination.

In the 1988 multi-wavelength campaign it was argued that the UV flux increased from
HB - NB - FB (Vrtilek et al. 1990). As the UV flux was regarded as a good measure of a mass 
accretion rate $\dot M$, it was proposed that $\dot M$ increases monotonically from HB - NB - FB, 
i.e. suggesting that the physical changes between the three states in the Z-track sources in 
general are driven by mass accretion rate (Hasinger et al. 1990). This became adopted generally 
as a standard model; 
however as realized at the time by Hasinger and co-workers the {\it decrease} of X-ray intensity 
on the NB between Hard and Soft Apex runs counter to simple expectations. Moreover, it is clear
from the present work that the Flaring Branch in Cygnus\th X-2 in colour-colour diagrams actually
consists of absorption dips while real flaring data is only occasionally seen.

\begin{figure}[ht!]
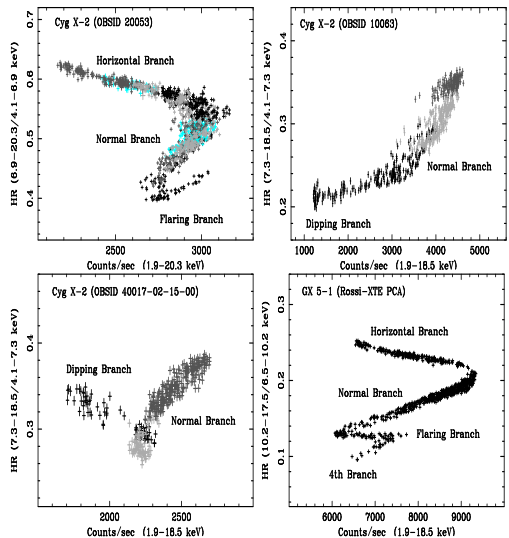
                                                                
\begin{flushleft}
\includegraphics[width=35mm,height=32mm,angle=270]{balucinska_2011_01_fig01.ps} 
\hskip 0.5 mm
\includegraphics[width=35mm,height=32mm,angle=270]{balucinska_2011_01_fig02.ps}    
\includegraphics[width=35mm,height=32mm,angle=270]{balucinska_2011_01_fig03.ps}
\hskip 1 mm
\includegraphics[width=35mm,height=32mm,angle=270]{balucinska_2011_01_fig04.ps} 
\caption{\footnotesize A variety of Z-track effects in the Cyg-like sources shown in
hardness-intensity, including intensity reductions in dipping.}
\label{Z-tracks}
\end{flushleft}
\end{figure}

\section{The Dipping/Flaring confusion}

It has been known for some time that the Cyg-like sources show extra tracks in hardness-intensity
addition to the three main branches. In Fig. 1, four examples are shown using {\it Rossi-XTE} data. 
Firstly, we show a normal Z-track for Cygnus\th X-2 (top left panel) with HB, NB and FB. In fact, based
on examining all of the data in the {\it RXTE} archive, it is found that a normal Flaring Branch
occurs rarely in this source. The next two panels also show observations of Cygnus\th X-2
not executing a full Z-track, but displaying definite intensity reductions from about the
location of the soft apex between NB and FB. Such decreases are seen in many 
observations and are discussed further below. Finally, in the lower panel, 
a Z-track is shown for GX\th 5-1 with a fourth branch consisting of intensity 
reductions from the peak of flaring. Similar behaviour is seen in 
GX\th 340+0 clearly also suggestive of absorption. 

\begin{figure}[ht!]
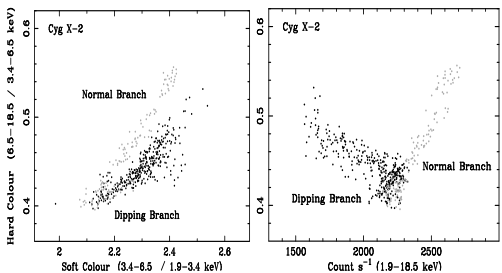
                                                              
\begin{flushleft}
\includegraphics[width=35mm,height=32mm,angle=270]{balucinska_2011_01_fig05.ps} 
\includegraphics[width=35mm,height=32mm,angle=270]{balucinska_2011_01_fig06.ps}  
\caption{\footnotesize Hardness-intensity and colour-colour diagrams of {\it Rossi-XTE} data showing
intensity decreases (dips) in Cyg\th X-2.} 
\label{CCD_HID}
\end{flushleft}
\end{figure}

In Fig. 2, we show intensity decreases in {\it RXTE} archival data on 
Cygnus\th X-2 in the two forms: hardness-intensity and colour-colour. While it 
is clear from hardness-intensity that 
\begin{figure*}[ht!]                                                 
\begin{center}
\includegraphics[width=50mm,height=140mm,angle=270]{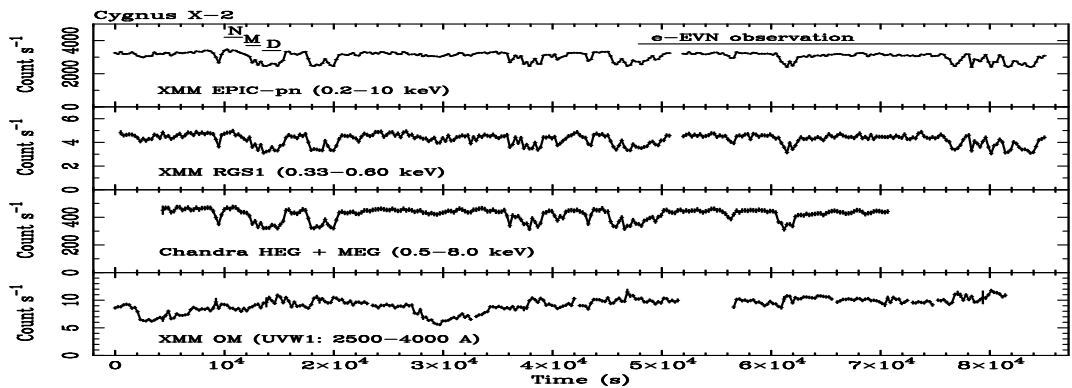}  
\caption{\footnotesize Lightcurves of the {\it XMM} EPC-pn and RGS instruments, the {\it Chandra} HETGS 
and the {\it XMM} Optical Monitor from the 2009 multi-wavelength campaign on Cyg\th  X-2.}
\label{}
\end{center}
\end{figure*}
\begin{figure*}[ht!]                                                  
\begin{center}
\includegraphics[width=50mm,height=140mm,angle=270]{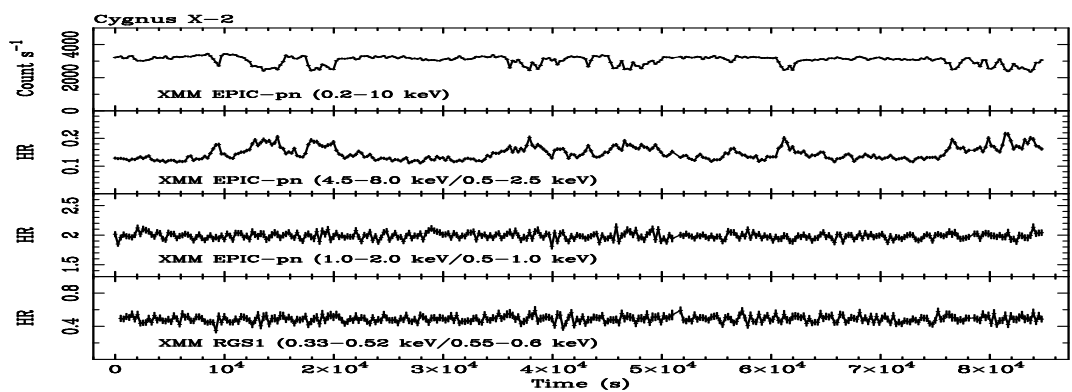} 
\caption{\footnotesize Hardness ratios from the observations (see text); also shown for comparison is the {\it XMM}
EPIC-pn lightcurve (top panel).}
\label{}
\end{center}
\end{figure*}
intensity decreases take place
visible as dips in the lightcurve, when plotted in colour-colour these events 
appear to look like flaring, demonstrating the nature of the confusion.

\section {Resolving the dipping controversy in Cygnus\th X-2: the multi-wavelength campaign of 2009}

We observed Cygnus\th X-2 in a multi-wavelength campaign in 2009, lasting 
24 hours from May 12, 9:30 UT to  May 13, 9:20 UT using {\it Newton XMM} and 
{\it Chandra} simultaneously and the European VLBI network at 5 GHz (Ba\l uci\'nska-Church et al. 2011). 
In addition UV data in the band 2500 -  4000 \AA\  were provided by the {\it XMM} Optical Monitor. 
X-ray data were obtained from the {\it XMM} EPIC-pn
camera in the band 0.6 - 12 keV using burst mode because of the brightness of the source
and the RGS instrument providing high resolution spectra in wavelengths 20.7-37.6 \AA.
The {\it Chandra} MEG and HEG and gratings provided high resolution spectra at
wavelengths greater than $\sim$ 13 \AA\  (0.5 - 8.0 keV).

Lightcurves of the observation in the X-ray instruments: the EPIC-pn, the RGS and the {\it Chandra}
MEG and HEG are shown in Fig. 3 together with the Optical Monitor lightcurve. In the top panel,
the second part of the observation covered by radio observations is indicated by a bar; the radio
flux was less than 150 microJy/beam, i.e. the source was radio quiet (Rushton et al. 2009) consistent
with the source's position on the Z-track (below) away from the Hard Apex. The X-ray data exhibit
extensive dipping as seen in Fig. 3.
In Fig. 4 we show a hardness ratio as a function of time for the EPIC-pn defined as the ratio
of counts per second in the band 4.5 - 8.0 keV to that in 0.5 - 2.5 keV. A clear increase in
hardness ratio is seen at the intensity dips suggestive of photoelectric absorption. However,
remarkably, hardness ratios based on an energy band below 2 keV in EPIC and a similar ratio using RGS data
showed no change in hardness in dipping as will be discussed later.

From these data, the evolution of Cygnus\th X-2 in hardness-intensity was derived as shown in 
Fig. 5. The source did not execute movement along the Z-track but remained at a stable position
throughout the 24 hours of observation. Spectral fitting (below) 
\begin{figure}[ht!]                                             
\begin{center}
\includegraphics[width=44mm,height=50mm,angle=270]{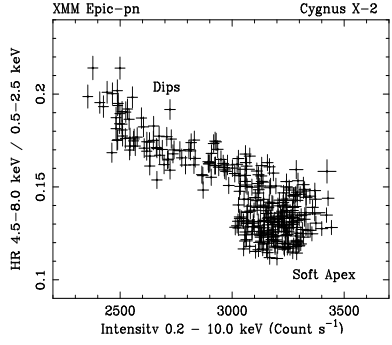} 
\caption{\footnotesize Hardness-intensity variation of the {\it XMM} EPIC-pn Cygnus\th X-2 data.}
\label{}
\end{center}
\end{figure}
reveals parameter values 
expected at or close to the Soft Apex. The excursions to lower intensity were found to correspond
to dips in the light curve.

\subsection{Continuum spectral fitting}

Non-dip and dip spectra were fitted using the Extended Accretion Disc Corona model 
(Church \& Ba\l uci\'nska-Church 2004) as justified by the now very strong evidence 
for an extended ADC (Ba\l uci\'nska-Church et al. 2011). The emission is dominated
as in LMXB in general by Comptonized emission and this model recognizes that for
an ADC extending to large radial distances, the seed photons will consist of
thermal emission from the accretion disk. Thus a model was used consisting of blackbody 
emission from the neutron star plus a cut-off power law Comptonized emission.

Fitting non-dip, intermediate and deep dip spectra showed that dipping consisted of absorption 
of partially covered Comptonized emission. However, remarkably, emission from the neutron star
was not absorbed at all, contrasting radically with dipping in the dipping class of  
LMXB (e.g. Church et al. 1997). Best fit models and unfolded data are shown in Fig. 6;
The decrease of the Comptonization due to photoelectric absorption is clear, and of an emission 
feature at $\sim$1 keV. However, the lack of change of the broad neutron star blackbody peaking
at 3 keV is also very clear, showing that the neutron star is not overlapped by absorber.
\begin{figure}[]                                          
\begin{center}
\includegraphics[width=44mm,height=50mm,angle=270]{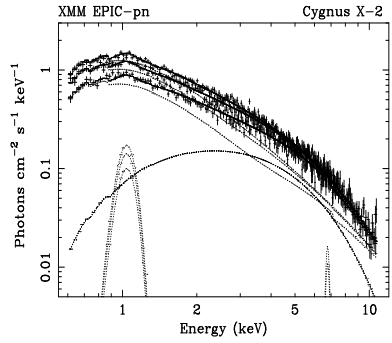} 
\caption{\footnotesize Spectral fitting of Cyg X-2 {\it XMM} EPIC non-dip, intermediate and deep dip spectra.}
\label{}
\end{center}
\end{figure}

\subsection{The grating spectra}

In Fig. 7 we show the {\it XMM} RGS grating spectrum of non-dip data revealing a strong absorption feature 
identified as the Oxygen edge. The {\it Chandra} HEGTS spectra 
reveal a number of absorption edges of Mg K, Ne K, Fe L and O K at 9.48 \AA\  (1308 eV),
14.29 \AA\ (867 eV), 17.51 \AA\ (708 eV) and 22.89 \AA\ (542 eV). 
Fig. 8 shows non-dip data (upper panel) and dip data (lower panel).
Remarkably, investigation of the edges
notably the Ne K edge in the {\it Chandra} MEG and the oxygen edge in the RGS showed the optical depth 
did not change in dipping. This can be explained in terms of a partial covering model.

\begin{figure}[]                                             
\begin{center}
\includegraphics[width=50mm,height=50mm,angle=0]{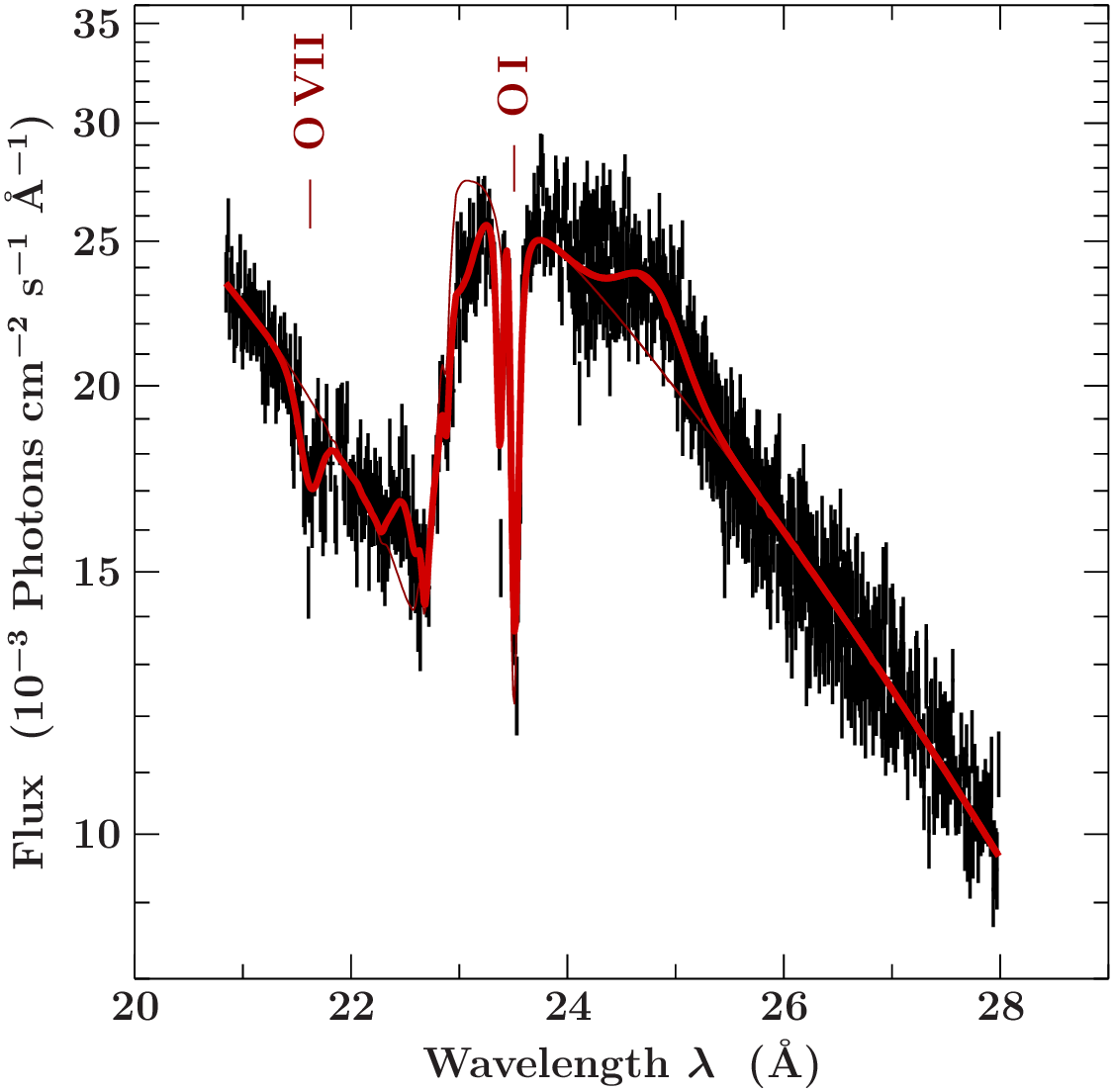} 
\caption{\footnotesize First order RGS spectrum of Cyg X-2 in the region of the O-edge, together with the best fit model.}
\label{}
\end{center}
\end{figure}
\begin{figure}[]                                                   
\begin{center}
\includegraphics[width=50mm,height=60mm,angle=0]{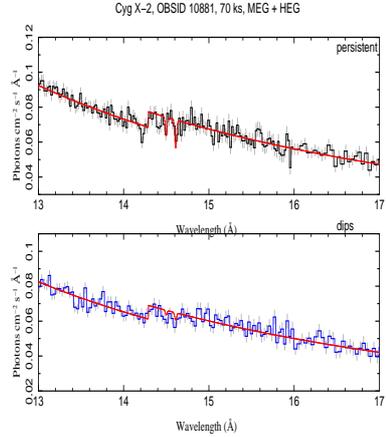} 
\caption{\footnotesize {\it Chandra} high resolution non-dip (upper panel) and dip spectrum
(lower panel) of Cyg X-2  in the vicinity of the Ne K edge.}
\label{}
\end{center}
\end{figure}
\subsection{Model for dipping in Cygnus\th X-2}

Dipping is unusual in this source in several respects: the neutron star emission is not absorbed;
the optical depth in the edges does not increase in dipping and the hardness ratios defined below
2 keV do not change in dipping (Fig. 4). These features can be explained as follows.
The major effect in dipping is a gradual part removal of the Comptonized emission.
Gradual removal is well-modelled by partial covering and proves the extended nature of the ADC
emission. Given the high inclination of the source of 62$\degmark$ it is clear that the absorber will
be structure in the outer accretion disk. The spectra are well described with this model: in 
intermediate dipping with a covering factor of 18\% and column density \NH = $13 \times 10^{22}$ atom cm$^{-2}$ 
and in deep dipping with a covering factor of 42\% and \NH = $45 \times 10^{22}$ atom cm$^{-2}$.
The contributions of uncovered and covered emission to the continuum in deep dipping are shown in
Fig. 9 using the continuum fit results (neglecting lines and edges).
\begin{figure}[h!]                                                            
\begin{center}
\includegraphics[width=45mm,height=44mm,angle=270]{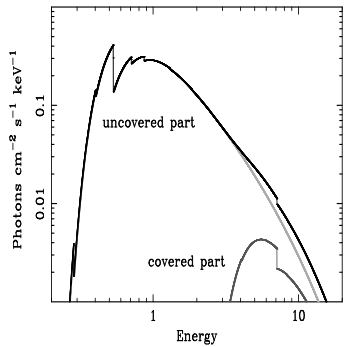} 
\caption{\footnotesize Best-fit to the deep dip spectrum showing the covered and uncovered
components demonstrating the total lack of flux of covered emission at low energies.}
\label{}
\end{center}
\end{figure}
The neutron star is not overlapped by absorber, totally different from the LMXB dipping sources
(e.g. XB\th 1916-053, Church et al. 1997) where the neutron star is always absorbed. However, this can
be explained in terms of the inclination not being quite so high as in the dipping LMXB; thus
the  envelope of the absorber in the outer disk covers a fraction of the ADC but misses the neutron star.
Because of the high column densities in deep dipping, covered emission is totally removed
below $\sim $ 4 keV. Thus the hardness ratios define below 2 keV do not contain covered, only
uncovered emission and so do not change in dipping. Similarly, the remarkable result that the
grating data show no increase of optical depth in dipping is due to the fact that only uncovered
emission is seen in dipping. Thus the gratings reveal the decrease of continuum intensity
but cannot see any of the absorbed emission which would have higher optical depth.

\subsection{Dependence of dipping on orbital phase}

The multi-wavelength observation covered orbital phase of Cyg X-2 between 0.32 and 0.42.
In the dipping LMXB dipping is normally confined to phase $\sim$0.75 corresponding
to the bulge in the outer disk. Clearly in the present data absorption takes place on the opposite
side of the disk. To investigate further we have derived orbital phases of dipping for all observations
of Cygnus\th X-2 containing dipping in the {\it RXTE} archive. Examination of all the {\it RXTE} ASM data 
in the archive as shown in Fig. 10
revealed that dipping is very common in this source, i.e. in almost all observations and 
using pointed data have derived the distribution of dipping with orbital phase as shown in Fig. 11.
In this figure, the main peak of the distribution occurs at phase $\sim$0.75 indicative of
absorption in the bulge in the outer disk where the accretion stream impacts. However, dipping takes 
place at all orbital phases suggesting structure around the accretion disk rim. Thus, in conclusion
there is no doubt that the Dipping Branch seen in the hardness-intensity diagrams of Cygnus\th X-2
consists of photoelectric absorption and is unrelated to flaring.

\begin{figure*}[ht!]                                                      
\begin{center}
\includegraphics[width=44mm,height=140mm,angle=270]{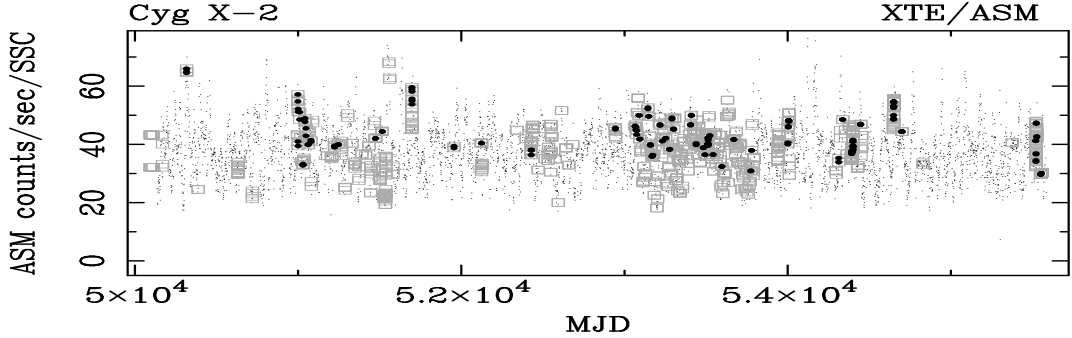} 
\caption{\footnotesize Cygnus\th X-2 lightcurve from the 15-year {\it RXTE} ASM archive;
intensity dipping (black circles) was identified using pointed observations (open squares).}
\label{}
\end{center}
\end{figure*}
\section{Dipping in GX\th 5-1 and GX\th 340+0}

The confusion between dipping and flaring mainly related to the source Cygnus\th X-2. However, 
in Fig. 1 we showed a Z-track in hardness-intensity for the Cyg\th X-2 like source GX\th 5-1 
in which a 4th track is seen of decreasing intensity attached to the end of the Flaring
Branch. The source GX\th 340+0 displays a similar 4th track. It might be supposed that
these 4th tracks are related to absorption because of the intensity decreases and so
are clearly relevant to the dipping/flaring confusion, so this effect is addressed here.

We have carried out spectral fitting of the evolution along the Z-track including the 4th track
(Church et al. 2010). This showed that there was no trace of absorption on the 4th track.
It was significant that spectral parameters such as the neutron star blackbody temperature 
$kT$ and blackbody radius $R_{\rm BB}$ did not exhibit change of behaviour when moving
from the Flaring Branch to the 4th Branch, but exhibited the same continuous smooth variations
on both branches indicating no marked physical change. However, the luminosity of the dominant
ADC Comptonized emission, essentially constant on the FB, exhibited a decrease on the 4th Branch 
as also seen on the NB between the Hard and Soft Apex. This strongly suggests a decrease
of mass accretion rate and shows that the 4th Branch consists of a continuation of $\dot M$
decrease which was already taking place as the source moved down the Normal Branch to the
Soft Apex at which point flaring began, identified as unstable nuclear burning (Church et al. 2006).
The 4th Branch thus consists of the continuation of decrease of $\dot M$ during the flaring
and the significant result is that it is unconnected with absorption.

\begin{figure}[]                                                       
\begin{center}
\includegraphics[width=44mm,height=50mm,angle=270]{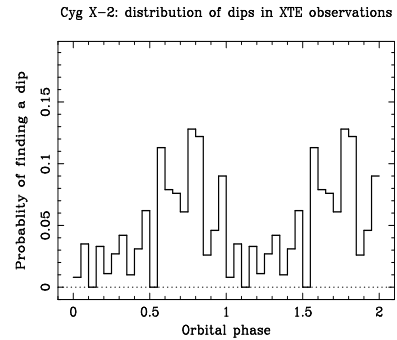} 
\caption{\footnotesize Probability of finding a dip in Cyg X-2 as a function of orbital phase.}
\label{norm_hist_dips}
\end{center}
\end{figure}

\section {Conclusions}

Based on a multi-wavelength campaign on Cygnus\th X-2 we have resolved the dipping/flaring 
confusion in this source. In these observations, strong dipping took place and spectral analysis
showed this to consist of photoelectric absorption in which $\sim$40\% of the extended Comptonized
emission of the ADC was overlapped by absorber in the outer disk; however, the neutron star
emission was not absorbed indicating that it was not overlapped by absorber. The grating spectrometers
also showed the dips as continuum decreases but without increase of optical depth in dipping
as all emission below 4 keV was removed.

The original argument that dip data displayed in a colour-colour diagram appeared
very similar to a Flaring Branch led to the idea that the mass accretion rate increased
on this branch and  that this increased monotonically around the Z-track
in the direction HB - NB - FB. Moreover the detection of dipping in Cyg\th X-2 and GX\th 340+0
led to the idea that this sub-group of Z-track
sources differed from the Sco\th X-1 like sources by having high inclination. While dipping
clearly takes place in Cygnus\th X-2 having high inclination, we now see that apparent dipping
in {\hbox {GX\th 5-1}} and GX\th 340+0 probably relates to the 4th Branch and is not absorption,
so the motivation for inclination  distinguishing Cyg-like and Sco-like sources
no longer remains. Similarly, the motivation for $\dot M$ increasing in the 
direction HB - NB - FB does not remain. 

We have previously suggested a physical model for the Cyg-like
sources (Ba\l uci\'nska-Church et al. 2010) based on increase of mass accretion on the NB and unstable nuclear burning
on the FB. We have also recently proposed an explanation of the Sco-like source behaviour
in Sco\th X-1, GX\th 349+2 and GX\th 17+2 (Church \& Ba\l uci\'nska-Church (2011: {\it these
Proceedings}). In this explanation, the sources are dominated by almost non-stop flaring
which releases energy on the neutron star leading to much higher neutron star temperatures:
the main spectral difference.

In the other Cyg-like sources GX\th 340+0 and GX\th 5-1, the 4th Branch in which the X-ray intensity
decreases from the end of the Flaring Branch has been shown to be unrelated to absorption
dipping (Church et al. 2010).

Finally, the investigation of dipping in Cygnus\th X-2 provides additional proof of the
extended nature of the Comptonizing ADC, adding to the proof provided by the technique
of dip ingress timing (Church \& Ba\l uci\'nska-Church 204) and by the Doppler widths
of highly ionized emission lines (Schulz et al. 2009).
The gradual removal of this emission 
as clearly seen in the development of dipping, and formalized by fitting of a partial
covering model with increasing covering fraction, would not be possible if the Comptonizing
region was not extended. Thus the conception of the Comptonizing region widely held by 
parts of the community as a small, hot central region (by definition a point source
given a size of 1000 km or less) is no longer tenable.

\begin{acknowledgements}
This work was supported in part by the Polish Ministry of 
Science and Higher Education grant 3946/B/H03/2008/34.
\end{acknowledgements}

\bibliographystyle{aa}

\bigskip
\bigskip
\noindent {\bf DISCUSSION}

\bigskip
\noindent {\bf WOLFGANG KUNDT:} I was impressed by your careful treatment of these bright
low mass X-ray binary sources. Where may I find a concise description of your results ? 

\bigskip
\noindent {\bf MONIKA BA\L UCI\'NSKA-CHURCH:} The recent results are described in 
Ba\l uci\'nska-Church et al. (2011) and Church \& Ba\l uci\'nska-Church ({\it these Proceedings}).

\end{document}